# Application of Graph Based Vision Transformers Architectures for Accurate Temperature Prediction in Fiber Specklegram Sensors.

Abhishek Sebastian[1,2]

[1] University of Sydney
[2] Division of Applied AI, Abhira
`Contactabhisheksebastian@gmail.com`

**Abstract.** Fiber Specklegram Sensors (FSS) are highly effective for environmental monitoring, particularly for detecting temperature changes; however, the nonlinear nature of specklegram data presents significant challenges for accurate temperature prediction. This study explores the application of transformer-based architectures, such as Vision Transformers (ViTs), Swin Transformers, and novel models like Learnable Importance Non-Symmetric Attention Vision Transformers (LINA-ViT) and Multi-adaptive Proximity Vision Graph Attention Transformers (MAP-ViGAT), to predict temperature using specklegram data over a range of 0–120°C. The results show that ViTs achieved the lowest Mean Absolute Error (MAE) of 1.15, outperforming traditional models, such as CNNs. GAT-ViT and MAP-ViGAT variants demonstrated competitive accuracy, emphasizing the significance of adaptive attention mechanisms and graph-based structures in capturing intricate modal interactions and phase shifts in specklegram data. Additionally, this study incorporates Explainable AI (XAI) methods, including attention and saliency maps, to provide insights into the decision-making process of ViT models for specklegrams, enhancing interpretability and transparency. These findings establish transformer architectures as new benchmarks for optical fiber-based temperature sensing, offering promising avenues for future applications in industrial monitoring and structural health assessments.

**Keywords:** Fiber Specklegram Sensors, Vision Transformers, Swin transformers, swin residual transformers, Learnable importance non-symmetric attention vision transformers, Multi-adaptive proximity vision attention transformers, graph attention networks, modal interference patterns, and explainable AI.

## 1 INTRODUCTION

Fiber Specklegram Sensors (FSS) have become essential tools for environmental monitoring because of their high sensitivity to temperature changes. By leveraging the interference patterns of coherent light propagating through multimode fibers, these sensors can detect subtle changes in the surrounding environment [31]. However, the complex and nonlinear nature of specklegram data presents significant challenges for the accurate prediction of temperature variation. Traditional machine learning models, such as Convolutional Neural Networks (CNNs), have shown promise but often fail to address the intricacies of the data [32], particularly when handling global phase relationships and intensity variations. This gap underscores the need for more advanced methodologies capable of capturing intricate dependencies present in specklegram images.

Although the scientific community has made considerable progress with deep learning models, there remains a growing interest in applying transformer-based architectures, which have demonstrated exceptional performance in natural language processing and image classification tasks. However, the challenge lies in adapting these models, such as Vision Transformers (ViTs) [21] and Swin Transformers [18], for temperature prediction in FSS systems. Existing research has primarily focused on traditional architectures and the potential of transformers in this domain remains largely unexplored. There is a need to move beyond conventional methods and develop models that not only achieve higher accuracy but also offer better interpretability and real-time applications in industrial environments[22][25][26][27].

To address this need, this study investigated the application of transformer-based models for temperature prediction using specklegram data. We present a comparative analysis of several transformer architectures, including ViTs, Swin Transformers, and two novel models: learnable-importance non-symmetric attention vision transformer (LINA-ViT) and multi-adaptive proximity vision graph attention transformer (MAP-ViGAT). These models were trained on a dataset of specklegrams captured over a wide temperature range and evaluated against traditional deep learning approaches to assess their efficacy. Additionally, we incorporate Explainable AI (XAI) techniques, such as attention and saliency maps, to provide insights into the decision-making processes of the models and enhance the interpretability of the results.

The remainder of this paper is organized as follows. Section 2 provides a comprehensive literature review of FSS technology and the evolution of vision transformer models for various data analyses. Section 3 describes the dataset and preprocessing techniques employed. Section 4 presents the methodology and architecture details of each model. Section 5 discusses the experimental results and compares the performance of transformer-based models with traditional methods. Finally, Section 6 concludes with insights into future research directions and the potential real-world applications of these advanced models.

## 2     Literature Survey

Fiber specklegram sensing (FSS) has emerged as a versatile and powerful optical sensing technology with applications spanning refractive index (RI) sensing, temperature measurement, water leak localization, torsion measurement, and pressure sensing. For instance, Al Zain et al. [1] introduced a highly sensitive multimode optical fiber (MMF) specklegram RI sensor incorporating a microfiber (MF) with a 33-μm waist diameter. This configuration demonstrated enhanced RI sensitivity, as quantified through zero-mean normalized cross-correlation (ZNCC) analysis, and exhibited temperature sensitivity of $0.013°C^{-1}$ across the range of 25°C to 65°C. Similarly, Wang et al. [5] explored temperature sensing using specklegram analysis, revealing a linear relationship between the correlation coefficients and temperature variations within an MMF-SMF splice system. In another application, Inalegwu et al. [2] utilized FSS for water leak detection, integrating convolutional neural networks (CNNs) to analyze specklegrams and achieving 100% accuracy in leak detection and localization, although experimental conditions caused slight accuracy reductions.

Li et al. [4] developed a torsion sensor using FSS and a residual network (ResNet) architecture, achieving a torsion prediction error of ±2° over a range of 0 –360°. The high accuracy of the model is attributed to effective specklegram decorrelation processes. Other studies have compared FSS-based temperature measurement systems with existing fiber optic technologies. For example, Vélez-Hoyos et al. [3] optimized an FSS for temperature sensing, with results comparable to those of Fiber Bragg Grating (FBG) sensors. In addition, Wang et al. [6] extended FSS to simultaneous temperature and weight measurements, showing a correlation between specklegram patterns and physical parameters, whereas Musin et al. [7] demonstrated a surface temperature sensor based on modal interference, emphasizing its advantages in spatially integrated temperature sensing.

Furthermore, Brestovacki et al. [8] introduced a low-cost Raspberry Pi-based system for specklegram image capture and analysis, which proved efficient for mechanical deformation sensing. Reja et al. [9] investigated the use of deep learning in pressure sensing, demonstrating that environmental crosstalk in specklegrams could be mitigated through a multilayer perceptron (MLP) model, achieving error rates below 3 kPa over a 1 MPa range. Collectively, this body of research underscores the adaptability of FSS technology across a broad range of sensing applications, particularly when combined with machine-learning techniques.

In parallel, Vision Transformers (ViTs) have emerged as robust alternatives to CNNs for computer vision tasks. Unlike CNNs, which rely on localized feature extraction through convolutional layers, ViTs leverage self-attention mechanisms to capture the global dependencies in images. Ovadia et al. [10] introduced the Vision Transformer-Operator (ViTO), which combines ViTs with operator learning for inverse PDE problem solving, showing their superiority over traditional operator networks in super-resolution tasks.

To address the limitations of ViTs in dense prediction tasks, Xia et al. [11] proposed ViT-CoMer, which incorporates multi-scale feature interactions to enhance spatial information exchange. Dehghani et al. [13] introduced NaViT, a model designed to handle arbitrary aspect ratios and resolutions, moving beyond the conventional resizing typically required by vision models.

Further advancements were made by Hatamizadeh and Kautz [14] with MambaVision, a hybrid architecture that integrates ViTs with Mamba blocks, enhancing their efficiency in visual feature modeling. To improve memory efficiency, Yun and Ro [15] proposed SHViT, which reduces computational redundancy by using a single-head attention mechanism. The application of ViTs has shown notable success across various domains, including thermal image super-resolution [17] and temperature forecasting [16], thereby highlighting their versatility.

The introduction of the Swin Transformer by Liu et al. [18] marked another significant advancement, demonstrating its potential as a general-purpose backbone. By employing shifted windows, the Swin Transformer efficiently manages both local and global information. Building on this, Yao and Shao [19] developed an ESwin Transformer, which enhances the performance on smaller datasets through redesigned patch embedding and merging modules.

Despite the success of transformer models in various computer vision tasks, their application to FSS for temperature monitoring remains largely unexplored. Traditional models have been thoroughly studied in this context; however, transformers present new opportunities for enhanced sensing capabilities.

In this study, we aim to address this gap by evaluating a range of pre-trained deep learning models originally designed for classification tasks on FSS data.

Vision Transformers (ViTs) and swin Transformers, both designed for classification on the ImageNet dataset, were re-engineered for regression tasks and trained on a specklegram image dataset to predict the temperature. In addition, we introduce two novel Vision Transformer architectures: LINA-ViT and MAP-ViGAT. LINA-ViT replaces standard multihead self-attention layers with LINA layers, incorporating a learnable importance mechanism into the non-symmetric attention process. By contrast, MAP-ViGAT leverages spatial, learnable, and feature adjacency matrices to graph patches, capturing intricate relationships between different regions of the specklegrams.

Although ViTs are known for their complexity [20][23][24], which can complicate interpretability, we addressed this challenge by applying a suite of interpretability tools to improve the transparency of specklegram assessments. This analysis offers new insights into the potential of transformers for FSS, paving the way for enhanced performance in optical sensing tasks.

## 3    MATERIALS AND METHODS

### 3.1    Description of the dataset used in the study

The dataset [30] comprises speckle patterns generated by modal interference in a multimode optical fiber. Modal interference occurs when multiple propagating modes within the fiber overlap and interfere owing to the differences in their phase velocities and propagation constants. These phase differences lead to constructive and destructive interference at various points across the fiber cross-section, resulting in speckle patterns observable at the fiber output end. Speckle patterns are highly sensitive to external perturbations, such as temperature changes, which affect both the refractive index (through the thermo-optic effect) and the physical dimensions of the fiber (due to thermal expansion). Specifically, temperature variations alter the effective refractive indices and propagation constants of the fiber modes, thereby changing the modal distribution and the resultant interference pattern.

The dataset features specklegrams that capture temperature-induced variations, simulated using the finite element method (FEM) to model the response of the fiber under thermal changes. The temperature was varied from 0 to 120 °C in increments of 0.2 °C, yielding 601 distinct specklegrams.

Each specklegram, with dimensions of 126 × 126 pixels, represents the intensity distribution at the output of the fiber, which captures the thermally induced changes in the interference patterns. These changes are inherently nonlinear because of the complex dependence of the modal propagation constants on the temperature and the subsequent modal interactions within the fiber.

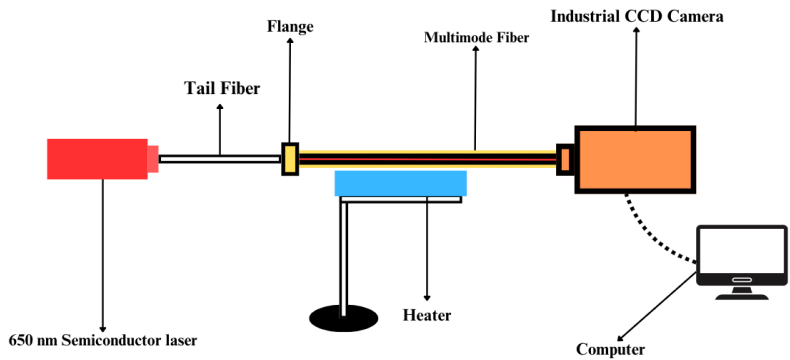

**Fig. 1. Schematic of the experimental setup for dataset collection.**

The experimental setup (illustrated in Figure 1) involves coupling a 650 nm semiconductor laser into a multimode fiber via a tail fiber and flange. As the laser light propagates through the multimode fiber, multiple guided modes are excited, each with its own propagation constant and spatial-field distribution. The superposition of these modes leads to interference effects that manifest as speckle patterns at the output end of the fiber. A heater was used to precisely regulate the temperature of the fiber, inducing controlled variations in the refractive index and physical dimensions of the fiber. These thermal effects cause changes in the effective indices and the phase accumulation of the modes, leading to shifts in the interference pattern. The output speckle patterns were captured using an industrial CCD camera with sufficient spatial resolution to resolve speckle features. These patterns were then analyzed computationally to study the effects of temperature-induced modal interference in the subsequent sections of this study.

### 3.2   Challenges in Analyzing Abstract Specklegram Images for Deep Learning

Analyzing specklegram images using deep learning models presents significant difficulties owing to the intricate and stochastic nature of the optical interference patterns they depict. Specklegrams are formed by coherent superposition of multiple optical modes propagating through a multimode fiber. Each mode has a distinct spatial field distribution and propagation constant, and their interference results in complex intensity patterns that are highly sensitive to environmental factors, such as temperature.

From an optical standpoint, temperature variations affect the refractive index of the fiber through the thermo-optic effect and alter its physical dimensions owing to thermal expansion. These changes affect the effective refractive indices and phase velocities of the modes, leading to shifts in the interference pattern. The resulting speckle patterns are granular and random, and lack distinct spatial features or structures. The intensity distribution follows a negative exponential probability density function characteristic of a fully developed speckle, embedding information in statistical and correlation properties rather than in recognizable shapes.

From a computer vision perspective, these specklegram images are difficult to interpret because they lack the clear, high-contrast edges or repetitive textures typically used by convolutional models to learn features. The intensity variations are governed by random phase differences between modes, causing the patterns to appear as fine-scale noise rather than coherent features. This randomness challenges the ability of deep neural networks to learn meaningful representations from data, as the models may not effectively capture the subtle statistical variations that correspond to changes in environmental parameters, such as temperature.

### 3.3 Data Preprocessing

Given the challenges posed by the abstract and stochastic nature of specklegram images, it is crucial to apply preprocessing techniques that effectively highlight the underlying features of deep-learning models. These Speckle patterns lack distinct spatial features, making it difficult for typical computer-vision approaches to extract meaningful information. To address this issue, we employed specialized preprocessing techniques designed to enhance feature extraction and capture the intricate intensity variations that are critical for accurate temperature prediction. These techniques aim to accentuate the specklegram's subtle textures and intensity gradients, thereby facilitating more effective learning in subsequent deep learning tasks.

**Local Binary Patterns (LBP).**

Local Binary Patterns (LBP) [28] are widely used texture descriptors in computer vision to characterize local spatial patterns within an image. The LBP algorithm operates by thresholding the neighborhood of each pixel and encoding the result into a binary code. Specifically, for each central pixel, the LBP compares its intensity with that of its surrounding neighbors within a defined radius. If a neighboring pixel has an intensity equal to or greater than the central pixel, it is assigned a "1"; otherwise, it is assigned a "0". These binary values are then concatenated to form a binary number (LBP code) for the central pixel. This process effectively captures local texture patterns by encoding the spatial relationships of pixels

In the context of specklegrams, however, random intensity variations due to optical interference mean that local intensity comparisons may not capture consistent or meaningful texture information related to the underlying physical phenomena. The stochastic nature of speckle patterns limits the effectiveness of LBP in extracting features that correlate with temperature-induced changes in the fiber.

**Gradient Computation.**

Gradient computation involves calculating the magnitude and direction of the intensity gradient at each pixel, which represents the rate of change in intensity [29]. In computer vision, gradients are fundamental for detecting edges and understanding structural information within images. They are typically computed using operators such as Sobel, Scharr, or Prewitt filters (in the current study, the Sobel filter is used), which approximates the first derivative of the intensity function in both the horizontal and vertical directions through convolution with specific kernels. The gradient magnitude highlights areas in the image where the intensity changes rapidly, corresponding to edges or boundaries.

For specklegram images, gradient computation can reveal the structural patterns associated with the interference of optical modes. The speckle pattern intensity varies owing to constructive and destructive interference, leading to regions of high and low intensities. Calculating the gradient emphasizes these regions of significant intensity change, effectively highlighting the fine-scale structures within the speckle pattern that correlate with temperature-induced variations in the fiber. By enhancing these subtle features, gradient images can improve the ability of deep-learning models to learn meaningful representations from speckle patterns.

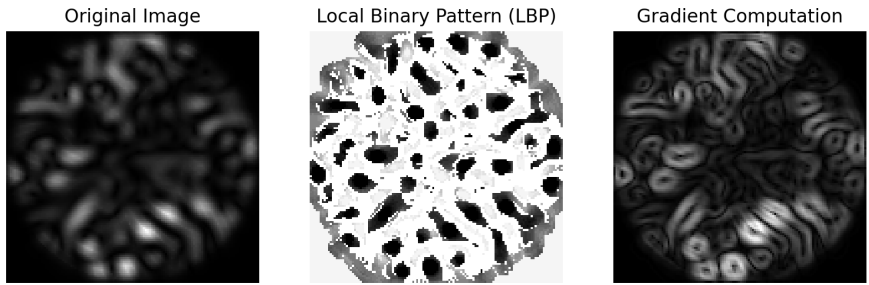

**Fig. 2.** Preprocessing steps were applied to the dataset. The figure illustrates the preprocessing stages, starting with the original image (left), followed by the Local Binary Pattern (LBP) representation (middle), and concluding with the gradient computation (right). These steps enhance the image features and aid in subsequent processing tasks.

We applied both LBP and gradient computations to the dataset to evaluate their effectiveness in enhancing the interpretability of specklegrams (see Figure 2). It was found that gradient computation provided a more substantial enhancement in feature representation than LBP. The gradient images accentuate the fine texture details and local intensity variations that are crucial for distinguishing subtle changes in speckle patterns resulting from temperature variations. This enhancement aligns with the optical understanding that temperature changes affect modal interference within the fiber, leading to alterations in the intensity gradients of the speckle patterns.

Consequently, gradient computation, along with normalization and resizing, was employed in the subsequent analysis to better reveal the underlying patterns in the specklegrams and improve the performance of the deep learning models. This approach bridges the gap between the optical properties of specklegrams and the requirements of computer-vision algorithms.

## 4      PROPOSED METHODOLOGY

In the methodology outlined in Figure 3, we employed a preprocessed dataset to train both the pre-trained models and custom architectures. We applied various quantitative analysis techniques such as MSE, MAE, RMSE, Maximum Error, and $R^2$ score to validate the accuracy and dependability of our results. The model training process was consistent throughout the study, with each model trained for 100 epochs using a 70:20:10 split for the training, testing, and validation sets.

## 4.1 Pre-trained models and their intermediate results.

**MobileNet pre-trained network.**

MobileNet, although originally selected for its computational efficiency in terms of speed and memory, encounters significant limitations when applied to fiber specklegram sensor (FSS) data. MobileNet, designed for large-scale image classification tasks, is optimized to capture distinct, high-level object features, making it less suited for handling fine, continuous variations observed in specklegrams.. The model's architecture, which employs depth-wise separable convolutions to reduce computational cost and parameter count, lacks the capacity to fully capture the high-dimensional, nuanced features in specklegram data. This is particularly problematic, given that FSS data involve intricate spatial and modal interactions that are sensitive to even minute environmental changes.

Despite adjustments to hyperparameters such as the learning rate, batch size, and dropout rate, MobileNet struggled to generalize well on specklegram data because of the inherent complexity and high spatial frequency information within these images. These limitations emphasize the need for more tailored architectures, possibly leveraging attention mechanisms or spectral-domain processing, which can more effectively handle the unique optical and modal characteristics of specklegram data, enabling a more accurate interpretation of temperature or strain-induced variations in the sensor.

**Inception Net V2 pre-trained network.**

Inception V2, although widely recognized for its ability to capture multi-scale features in complex images, showed significant limitations when applied to fiber specklegram sensor (FSS) data. Designed primarily for image classification tasks, Inception V2's architecture, which features parallel convolutional layers with varying kernel sizes, was found to struggle with modeling fine-grained, continuous interference patterns typical of specklegrams.

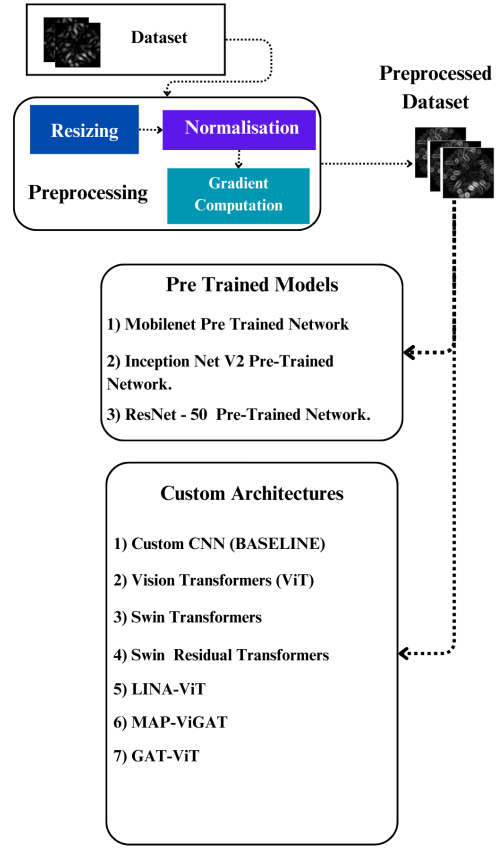

**Fig. 3.** Proposed Evaluation Workflow. This figure shows the workflow of the proposed evaluation process. The dataset underwent preprocessing, including resizing, normalization, and feature extraction. The processed data were then utilized for training two sets of models: Custom Architectures and Pre-trained Models.

Although Inception V2 is effective at extracting features across multiple scales, it proved insufficient for capturing the intricate, high-dimensional data associated with temperature changes in FSS images. The shortcomings of the model became evident through its higher Mean Absolute Error (MAE) and Mean Squared Error (MSE) during temperature prediction tasks (as detailed in Table 1). These performance metrics suggest that despite its capacity to handle diverse feature scales, Inception V2 struggles with the specialized demands of specklegram analysis.

**ResNet pre-trained network.**

ResNet, or Residual Networks, was selected due to its ability to handle the deep and complex nature of fiber specklegram sensor (FSS) data. The architecture's use of residual connections addresses the vanishing gradient issue, enabling the training of deep networks that can extract intricate high-dimensional features from specklegram images.

In this study, pre-trained ResNet models, such as ResNet-50, were fine-tuned to improve the performance of the FSS dataset. While the ResNet models outperformed MobileNet and Inception V2 in key metrics, including the Mean Absolute Error (MAE) and Mean Squared Error (MSE) (Table 1), the results still showed deviations that highlighted the challenges of achieving the high precision required for accurate temperature predictions in specklegram-based sensing applications.

**Table 1.** Evaluation Metrics of Pre-trained Models

| Model Type | MSE | MAE | RMSE | Maximum error | $R^2$ Score |
|---|---|---|---|---|---|
| Mobile Net | 168.5 | 8.78 | 12.98 | 38.58 | 0.848 |
| Inception Net V2 | 185.83 | 11.24 | 13.63 | 39.08 | 0.833 |
| Resnet | 84.62 | 7.03 | 9.2 | 26.16 | 0.924 |

The results presented in Table 1 indicate that ResNet's residual connections significantly improve the model's ability to capture intricate, nonlinear relationships between speckle patterns and temperature variations. This architectural advantage enhances both gradient flow and deep feature extraction, leading to superior performance compared with MobileNet and Inception V2. These findings underscore the value of residual connections in optimizing models for high-precision temperature measurements in FSS applications. The subsequent sections explore the potential of transformer models incorporating similar residual connections to further improve the accuracy and efficiency of temperature prediction models.

**Custom Architectures.**

**Convolutional neural network.**

In this study, a custom CNN was developed to process FSS data. Despite being less complex than advanced architectures, such as ResNet or Inception V2, this CNN was specifically tailored to address the unique optical characteristics of specklegram data and engineered to capture detailed interference patterns. The network starts with an input layer for specklegrams and progresses through convolutional layers with increasingly larger filter counts (64, 256, 192, and 128) to extract the hierarchical features. Rectified Linear Unit (ReLU) activations follow each convolutional layer, whereas max-pooling operations reduce spatial dimensions and retain crucial features. A dense layer with 2048 neurons aggregates these features, and dropout is implemented to mitigate overfitting. The output layer consists of a single neuron with a linear activation function for temperature prediction. Optimized using the Mean Squared Error (MSE) loss function and Adam optimizer, this architecture balances the model complexity and generalization, establishing a robust baseline for handling specklegram data.

**Vision Transformers (ViTs).**

The Vision Transformer (ViT) model [12], adapted from NLP transformers for image tasks, was designed to address the challenges of the FSS data. Instead of using a transfer learning approach with the pre-trained ViT model from Google (ViT-base-patch16-224) [21], which was originally designed for classification tasks, the model described in this study was modified from scratch for regression (see Figure 4). This adaptation is crucial for accurately predicting the temperature based on subtle interference patterns in specklegrams, which arise from modal noise in multimode fibers.

The model processes the input specklegrams of size (126, 126, 3), which are then divided into $16 \times 16$-pixel patches via a Conv2D layer. This segmentation transforms the 2D specklegram into a sequence of patches analogous to words in NLP, allowing the transformer to process localized interference patterns effectively.

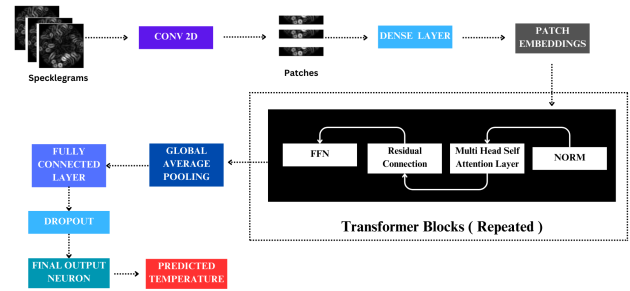

**Fig. 4.** Architecture of vision transformers for regression

These patches were then flattened and mapped to lower dimensional embeddings (64 dimensions). In addition, positional embeddings were included to retain the spatial information of the speckle patterns, which is vital for understanding the relative positions of the interference fringes within the specklegram.

The core of the model consists of multiple transformer blocks, each of which is responsible for learning the intricate relationships between patches. The number of transformer blocks was set to four (in our architecture), and each block contained the following components.

a. **Layer normalization and multihead self-attention:** This layer prepares data for multihead self-attention, which captures the dependencies between different patches of the specklegram. This mechanism is key for modeling the intricate nonlinear relationships between speckle patterns and temperature variations.

b. **Residual Connections:** These layers are used after the attention mechanism to address the vanishing gradient problem. They ensure stable learning by preserving the gradient flow through the deep network (similar to the residual networks discussed earlier), which is essential for capturing the complex modal interference patterns in specklegrams.

c. **Feed-Forward Network (FFN):** Each block features a feedforward network with two dense layers. The first layer expands the embedding dimensions (dimension is set to 128) to capture more complex features, whereas the second layer reduces them back to the original size, allowing the model to learn detailed transformations of the speckle patterns.

Post-transformer block layers include a global average pooling layer that consolidates the learned features from all patches into a single vector, summarizing the entire specklegram information. This vector was then passed through a fully connected layer with 2,048 neurons and ReLU activation, with dropout (0.5) applied to mitigate overfitting. The final output layer uses a linear activation function to predict the temperature.

**Swin transformers.**

The Swin Transformers discussed in this study are an advanced variant of Vision Transformers (ViTs) optimized for FSS applications. Unlike the original Swin Transformer models [18], which were designed for classification tasks by Liu et al. and trained on the ImageNet dataset, our adaptation was built from the ground up. This is essential because the training of the base model on ImageNet 1k may not align with the specific requirements of the specklegrams.

Swin Transformers incorporate hierarchical feature representations and local attention mechanisms, making them capable of analyzing FSS data. This design effectively balances local feature extraction with a global context understanding, which is crucial for addressing the complex modal interference patterns present in specklegrams.

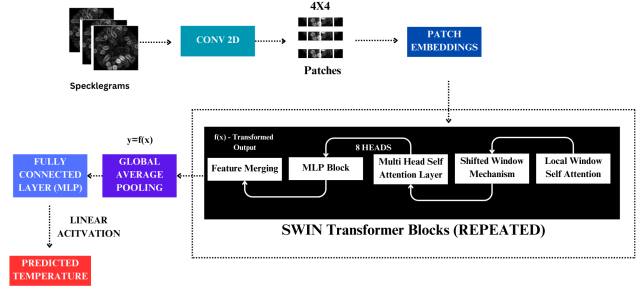

**Fig. 5.** Architecture of swin transformers for regression

Similar to Vision Transformers, Swin Transformers (see Figure 5) process input specklegrams of size (126, 126, and 3) and segment them into patches. However, instead of using global attention (as in ViTs), Swin Transformers apply self-attention within 4 × 4-pixel local windows. This approach reduces the computational complexity and enhances the capture of spatial dependencies between neighboring windows using a shifted window technique.

Each Swin Transformer block includes an MHSA mechanism (with six heads, compared to four in Vision Transformers) and a multilayer perceptron (MLP) block, but lacks residual connections. After processing through transformer blocks, a global average pooling layer consolidates the features into a single vector, which is then passed through a dense layer (with an MLP dimension of 512) to produce the final temperature prediction using a linear activation function.

Although Swin Transformers offer significant improvements in computational efficiency and local context integration through their hierarchical structure and window-based self-attention, they are still susceptible to the vanishing gradient problem owing to the absence of residual connections, unlike ViTs. To address this issue and stabilize the training process by allowing gradients to flow more easily through the network, residual connections are incorporated into the Swin Transformers in the next model discussed.

**Swin-residual transformers.**

Swin Residual Transformers, as presented in this study, combine the architecture of Swin Transformers with residual connections (see Figure 6) to enhance the model performance and stability for temperature prediction.

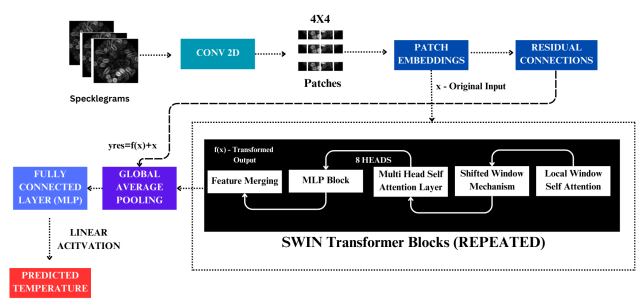

**Fig. 6.** Architecture of swin Residual transformers For Regression

The residual connections allow the model to learn the difference between the input and output of a layer. By combining the output of the final Swin Transformer block with the original input, the model preserves and integrates the initial optical embeddings, maintaining the intricate details of specklegram interference patterns.

The Swin Transformer (Figure 5.) performs a standard transformation of the input $x$ as shown in Equation 1.

$$y = f(x) \quad (1)$$

In Swin Residual Transformers, a residual connection modifies the output equation to

$$y_{res} = f(x) + x \quad (2)$$

where $y_{res}$ represents the output with the residual connection, $f(x)$ is the transformation by the Swin Transformer blocks, and $x$ is the original input.

Until now, the discussed architectures have shared a multihead self-attention layer. However, a major limitation is that the attention scores of patch embeddings are uniform or symmetric, meaning that every image patch contributes equally to the final output, which is an unrealistic scenario. To address this, the next model introduces a non-symmetric attention mechanism with learnable importance weights, resulting in a non-uniform distribution of importance across different patches in the image.

**Learnable Importance Non-Symmetric Attention Vision Transformer (LINA-ViT)**

**LINA-ViT**, introduced in this study, is derived from ViT and incorporates learnable importance weights and a non-symmetric attention mechanism (replacing traditional MHSA layers) to enhance the performance of complex data.

Similar to the other transformer architectures discussed earlier, the input and patching operations (see Figure 7) as well as the subsequent transformer blocks follow the same process.

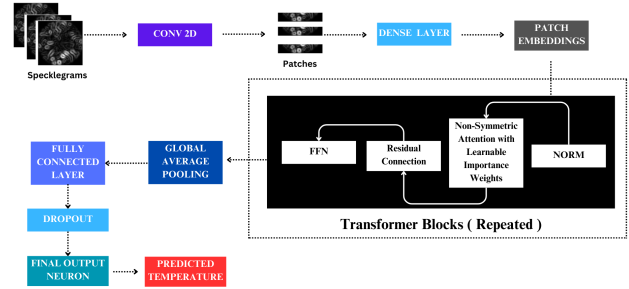

**Fig. 7.** Architecture of LINA-ViT For Regression

The key innovation in LINA-ViT is the involvement of a learnable importance weight layer. In this model, each patch of the input image was assigned a learnable scalar importance weight. This is achieved by introducing a layer that learns a unique importance weight for each patch during the training. These weights are initialized uniformly (typically set to 0) and fine-tuned as the training progresses depending on the task. The importance weights determine the relative significance of each patch, with the more important patches receiving higher weights. This allows the model to focus on patches that provide the most relevant information for a specific task while reducing the influence of less informative patches.

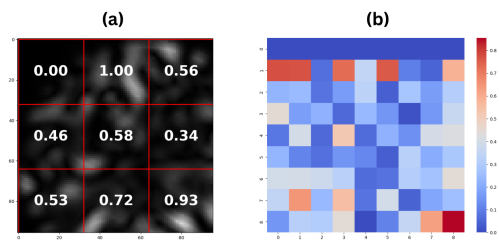

**Fig. 8.** **(a) Sample illustration showing how LINA-ViT assigns different importance weights to the image patches after training. The varying values represent the learned attention weights of each patch. (b) Sample illustration of the non-symmetric attention matrix generated by LINA-ViT, where the intensity of each cell indicates the attention score between different patches, highlighting the non-uniform distribution of importance across the image.**

During training, the model adjusts these weights through backpropagation, where the error gradient from the loss function (i.e., MSE for temperature prediction) updates the importance weights ( Figure 8). This ensures that the model learns to prioritize patches that carry more information regarding temperature-induced variations in the specklegram. Patches that contribute more to reducing the prediction error will increase their importance weights, enabling the model to focus more on these critical regions.

The non-symmetric attention (NSA) mechanism in LINA-ViT enhances the standard MHSA by incorporating learnable importance weights. Unlike traditional symmetric attention, in which all patches contribute equally, LINA-ViT modulates attention scores by scaling them with learned weights. This ensures that patches of greater significance have a higher impact on the final output.

This NSA mechanism is particularly crucial for tasks involving high-dimensional data such as specklegrams, where certain regions of the image may be more sensitive to environmental factors such as temperature. Patches representing these sensitive regions should receive more attention to ensure that the model accurately captures temperature variations, whereas patches with lower sensitivity can be downweighted. By dynamically adjusting attention using importance weights, LINA-ViT improves its ability to focus on the most relevant parts of the image, leading to better overall performance in complex tasks.

Building on LINA-ViT's use of learnable importance weights and non-symmetric attention, the next section explores the integration of Graph Attention Networks (GAT) without altering ViT's transformer blocks. GAT computes patch similarities before the transformer blocks, improving the efficiency over both ViT and LINA-ViT.

**Graph ATtention -Vision Transformer (GAT-ViT)**

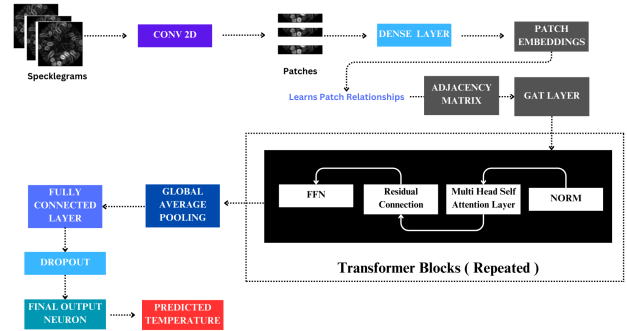

**Fig. 9.** Architecture of GAT-ViT For Regression

The GAT-ViT architecture combines Graph Attention Networks (GAT) with Vision Transformers (ViT) to model complex modal interference patterns in specklegrams for precise temperature predictions. GAT-ViT captures nonlinear dependencies between image patches, treating them as nodes in a fully connected graph, with the adjacency matrix reflecting the connectivity (see Figure 9). The attention weights highlight the most relevant interactions for an accurate prediction.

To further enhance the model's adaptability, we introduce the Multi-adaptive Proximity Vision Graph Attention Transformer (MAP-ViGAT). This advanced model dynamically constructs adjacency matrices, improving its ability to capture complex relationships within a specklegram.

**Multi-adaptive Proximity Vision Graph Attention Transformer (MAP- ViGAT)**

MAP-ViGAT (see Figure 10). introduced in this study was derived from the GAT-ViT. It shares foundational similarities with GAT-ViT, as both architectures combine Graph Attention Networks (GAT) with the ViT framework to capture complex relationships between image patches. However, a key distinction between GAT-ViT and MAP-ViGAT is how adjacency matrices are constructed to represent the relationships between patches.

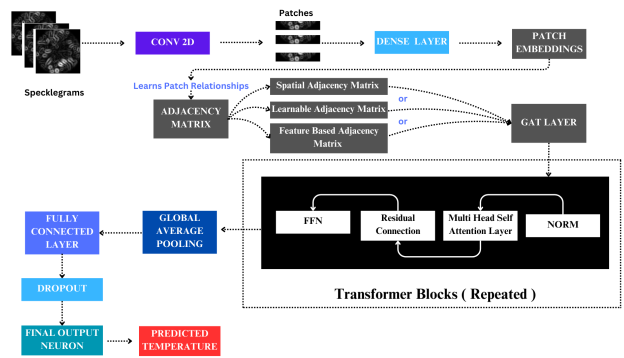

**Fig. 10.** Architecture of MAP-ViGAT For Regression

In GAT-ViT, the adjacency matrix is typically predefined and often represents a fully connected graph, implying that every patch is connected to every other patch. This approach is static and does not dynamically adapt to the specific characteristics of data.

In contrast, MAP-ViGAT introduces a multi-adaptive strategy for generating adjacency matrices (See Figures 11–13), offering more flexibility in defining the relationships between patches. Instead of relying on a single predefined graph structure, MAP-ViGAT dynamically constructs an adjacency matrix by using one of the following three strategies:

1.Spatial Adjacency Matrix (MAP-ViGAT (F)): This method emphasizes spatial proximity between patches based on their physical location in the image, often calculated using a Gaussian kernel to capture local interactions The relationship between patches $i\ and\ j$ is given by Equation 3.

$$A_{ij}^{(spatial)} = exp\left(-\frac{d_{ij}^2}{2\sigma^2}\right) \quad (3)$$

Where $d_{ij}^2$ is is the Euclidean distance between patches, and $\sigma$ controls the spread of the kernel.

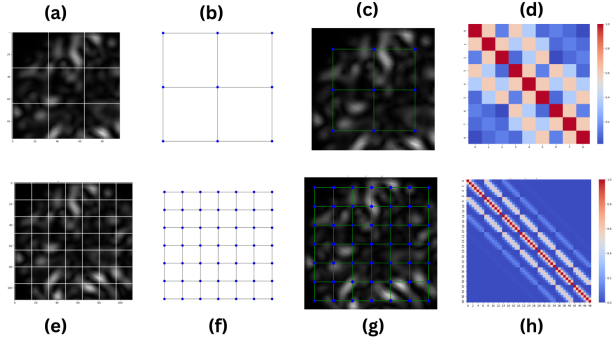

**Fig. 11.** (a) Patches of the sample specklegram (32 × 32).
(b) Creation of a graph using Graph Attention Networks (GAT) from the specklegram patches. (c) Graph overlaid on the specklegram showing connectivity between patches. (d) Spatial adjacency matrix corresponding to the specklegram patches (32x32). (e) Patches of the same sample specklegram (16x16). (f) Creation of a graph using GAT from the smaller specklegram patches. (g) Graph overlaid on the specklegram showing patch connectivity. (h) Spatial adjacency matrix corresponding to the specklegram patches (16x16)

**2.** Learnable Adjacency Matrix (MAP-ViGAT (L)): In this approach, the relationships between patches are learned dynamically during training (as shown in Equation 4.) This allows the network to adaptively discover the optimal structure based on specific characteristics of the input data.

$$A_{ij}^{(learnable)} = \theta_{ij} \quad (4)$$

where $\theta_{ij}$ are learnable parameters.

**3.** Feature-Based Adjacency Matrix (MAP-ViGAT (S)): This matrix is constructed based on the cosine similarity between patch embeddings (as in Equation 5.), capturing feature-based relationships where patches with similar content are more strongly connected, thus emphasizing global interactions across the image.

$$A_{ij}^{(feature)} = \frac{h_i \cdot h_j}{\|h_i\| \cdot \|h_j\|} \quad (5)$$

Where $h_i$ and $h_j$ are the feature vectors (embeddings) of patches *i* and *j*, and $\|.\|$ represents the vector norm.

Similar to other transformer architectures, the input and patching operations follow the same process ( See Figure 10).

Next, the patch embeddings are passed through a GAT Layer, which uses the adjacency matrix ($A$) to model the relationships between patches. GAT applies attention weights to the connections between patches, computed (as shown in Equation 6.).

$$\alpha_{ij} = \frac{\exp\left(ReLU\left(a^T[Wh_i||Wh_j]\right)\right)}{\sum_{k \in N(i)} \exp\left(ReLU\left((a^T[Wh_i||Wh_j]\right)\right)} \quad (6)$$

Where $\alpha_{ij}$ is the attention coefficient between patches $i$ and $j$, $a^T$ is a learnable attention vector, and $||$ denotes concatenation. $W$ is a weight matrix that transforms the input features.

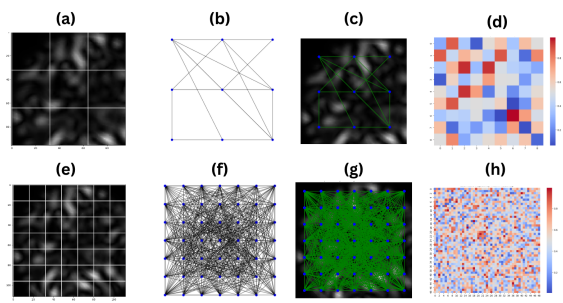

**Fig. 12. (a) Patches of the sample specklegram (32 × 32).**
**(b) Creation of a graph using Graph Attention Networks (GAT) from the specklegram patches. (c) Graph overlaid on the specklegram showing connectivity between patches. (d) learnable adjacency matrix corresponding to the specklegram patches (32x32). (e) Patches of the same sample specklegram (16x16). (f) Creation of a graph using GAT from the smaller specklegram patches. (g) Graph overlaid on the specklegram showing patch connectivity. (h) leanrable adjacency matrix corresponding to the specklegram patches (16x16)**

Other components of the models (GAT-ViT and MAP-ViGAT), such as the input, patching, and post-transformer blocks, maintain the same functionality as the previously discussed architectures (refer to ViT).

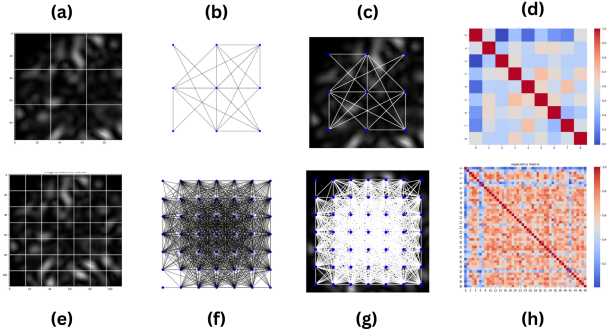

**Fig. 13.** (a) Patches of the sample specklegram (32 × 32).
(b) Creation of a graph using Graph Attention Networks (GAT) from specklegram patches. (c) Graph overlaid on the specklegram showing the connectivity between patches. (d) Feature-based adjacency matrix corresponding to specklegram patches (32 × 32). (e) Patches of the same sample specklegram (16 × 16 pixels). (f) Creation of a graph using GAT from smaller specklegram patches. (g) Graph overlaid on a specklegram showing patch connectivity. (h) Feature-based adjacency matrix corresponding to specklegram patches (16 × 16).

## 5   Visualization and explainability of Vision Transformer (ViT) with Attention Maps and Saliency Maps for Specklegrams.

In this section, we use two different tools to understand how Vision Transformers process specklegram data: attention and saliency maps. These tools provide insight into how the model interprets different regions of the specklegram during temperature estimation based on speckle patterns.

The attention map for every block and head is shown in **Figure 14**. Each row in the figure corresponds to the attention maps from a specific transformer block (from Block 1 to Block 4). Each column represents an attention map for a specific attention head within the block (from head 1 to head 4). The multihead attention mechanism allows the model to focus on different aspects of the input simultaneously, with each head learning different attention patterns. The color scale represents the magnitude of attention. The attention maps visualize the extent to which each patch in the input attends to every other patch. A darker color (purple) indicates lower attention, whereas brighter colors (yellow/green) represent higher attention values.

   The attention maps in Block 1 show very sparse attention with many isolated yellow points. This indicates that at the early stage of the model, certain patches focus on a few other regions but have not yet formed any broader patterns of dependencies. The sparse attention observed here could be because the model is just starting to parse the image, attempting to find the key regions of interest. Each head has a distinct focus, potentially on different areas of the image, meaning that they are learning to capture different low-level features such as edges or specific patterns in the input.

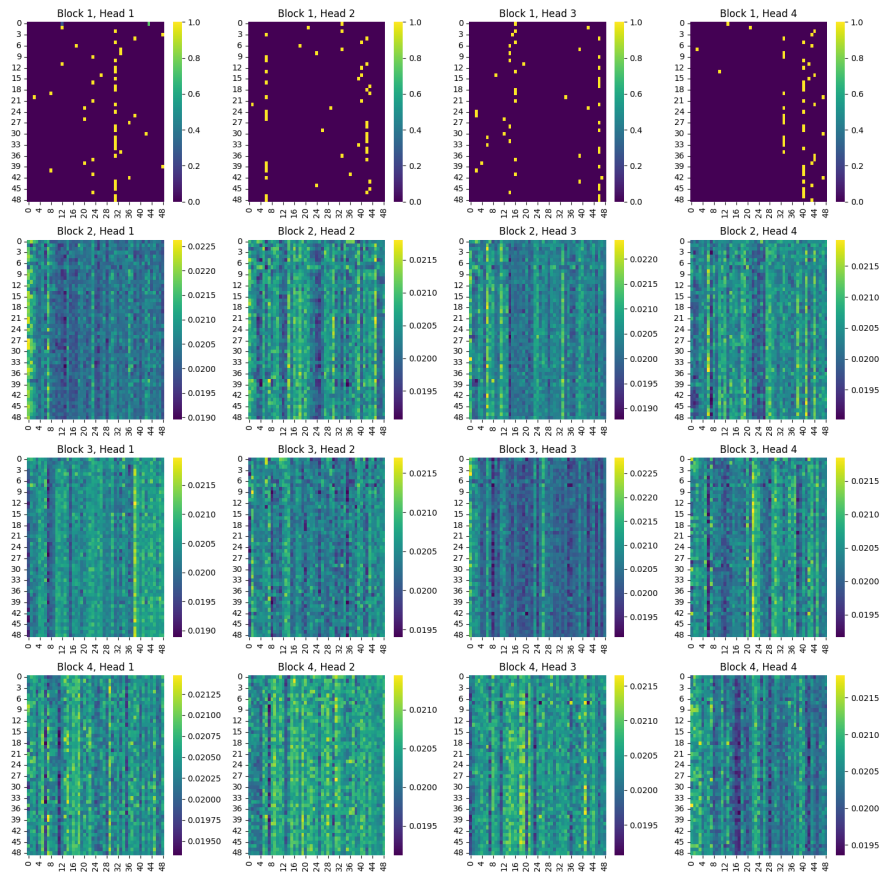

**Fig. 14.** Attention maps from Vision Transformers across four blocks and four heads illustrate how the model processes specklegram data.

In Block 2, the attention maps appear more structured than in Block 1, with some heads showing more defined vertical or horizontal streaks. This suggests that the model recognizes and focuses on broader and more consistent patterns within the input patches. The heads in Block 2 capture the relationships across wider regions of the image, indicating the emergence of mid-level feature representations (e.g., combining smaller local features into larger structures).

In Block 3, the attention maps exhibited even more refined patterns, with heads showing more defined bands or consistent streaks across the entire attention map. This suggests that the model captures more global dependencies across the different regions of the input image. As we move deeper into the transformer layers, the heads begin to focus on relationships across larger parts of the image, capturing more abstract features that involve interactions between distant regions of specklegram data. The emergence of these streaks indicates that the model identifies complex patterns spanning multiple patches, which are critical for temperature prediction in specklegram data.

By Block 4, the attention maps have become even more refined, with many heads showing distinct and well-distributed attention patterns. The model is likely to focus on the most important regions for making predictions, and has learned to attend to specific regions that are critical for understanding the input. This is the final stage before the predictions are made, and the heads concentrate on regions that contain the most relevant information for the task. The uniformity of attention in some cases indicates that the model aggregates global information to make accurate predictions.

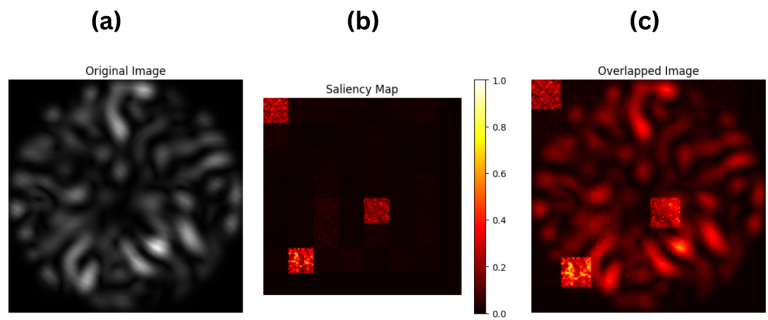

**Fig. 15. Saliency map analysis of specklegram data. (a) Original specklegram image input into Vision Transformer. (b) Generated saliency map, highlighting regions of the image that most influenced the model's prediction. (c) Saliency map overlaid on the original image; warmer colors (yellow/red) indicate higher importance, while cooler colors (black/blue) represent less significant areas. Saliency maps visualize the gradient of the model's output with respect to each pixel, helping interpret which parts of the image the model focuses on for prediction.**

Saliency maps ( Figure 15).  are visual representations of the regions in an image that are the most important or influential for a model's decision-making process. In ViTs, saliency maps highlight the areas on which the neural network focuses when making predictions. The goal is to visualize the contribution of each pixel of an input image to the final output, essentially helping to interpret which parts of the image the model "sees" as important.

When working with specklegrams, attention and saliency maps are crucial for understanding how deep learning models, such as Vision Transformers, process the data. Specklegrams that are sensitive to changes in surface texture, temperature, or scattering require careful analysis to extract meaningful patterns. Attention maps reveal how the model focuses on interference patterns or high-scattering regions such as areas of constructive interference. Saliency maps highlight the specific regions that most influence the model's decisions, offering insights into the key features used for tasks, such as temperature prediction.

# 6 RESULTS AND DISCUSSIONS

All models were rigorously evaluated using a standardized test dataset encompassing the full operational temperature range of 0–120 °C, as detailed in our methodology. The performance metrics are summarized in Tables 2 and 3. unequivocally demonstrated the superior capability of transformer-based architectures for extracting temperature-dependent information from FSS data. This section provides an in-depth analysis of the outcomes of the various models employed, their quantitative performance metrics, and a comparative discussion with existing literature in optical sensing

**Table 2.** Evaluation metrics for different custom architectures.

| Model | MSE | MAE | RMSE | Max error | $R^2$ Score |
|---|---|---|---|---|---|
| Custom CNN | 6.02 | 2.98 | 2.45 | 6.32 | 0.99 |
| Vision transformers | 1.95 | 1.15 | 1.4 | 3.67 | 0.99 |
| Swin Transformers | 98.83 | 7.27 | 9.94 | 34.05 | 0.91 |
| Swin Residual Transformers | 10.51 | 2.77 | 3.24 | 7.17 | 0.99 |
| GAT-ViT | 4.88 | 1.73 | 2.21 | 7.18 | 0.99 |
| MAP-ViGAT (F) | 9.56 | 2.27 | 3.09 | 9.93 | 0.99 |
| MAP-ViGAT (S) | 55.75 | 5.87 | 7.47 | 14.97 | 0.95 |
| MAP-ViGAT (L) | 26.59 | 3.35 | 5.16 | 40.87 | 0.97 |
| LINA-ViT | 16.43 | 2.93 | 4.05 | 15.96 | 0.98 |

Table 3. Comparison with other literary works.

| S.No | Literature | Evaluation Metrics | Architecture |
|------|------------|--------------------|--------------|
| 1 | Current Work | MAE: 1.15 | vision transformers |
| 2 | J D Arango et al [32] | MAE: 2.34 | CNN+ANN |
| 3 | Vélez H, F. et al [33] | MSE: 8.19 | ANN |

**Transformer Based Models**

ViT demonstrated the best performance, achieving a Mean Absolute Error (MAE) of 1.15, Root Mean Square Error (RMSE) of 1.4, and $R^2$ score of 0.99. These results emphasize the ability of ViT to effectively model both localized intensity variations and global phase relationships in specklegram patterns. By capturing the global dependencies and long-range interactions within the specklegrams, ViT outperformed traditional models, proving to be well suited for temperature prediction in FSS data.

Swin Transformer, on the other hand, exhibited suboptimal performance with an MAE of 7.27. Its hierarchical structure and shifted window approach, designed to capture multi-scale features in structured images, was ill-suited for unstructured and highly complex speckle patterns generated by multimode fiber interference. The Swin Transformer with residual connections improved the MAE to 2.77, showing that architectural modifications such as residual connections can enhance gradient flow during training, leading to better performance. However, the underlying architectural limitations of the Swin Transformer remain a challenge in effectively capturing global phase relationships in the FSS data.

The Graph Attention Vision Transformer (GAT-ViT) also performed well, with an MAE of 1.73 and an RMSE of 2.21. The graph-based approach allows GAT-ViT to capture the complex relationships between different regions of specklegram data. By treating image patches as nodes and using attention mechanisms on their connections, GAT-ViT effectively modeled local and nonlocal interactions, leading to high accuracy in temperature estimation.

MAP-ViGAT, with its various adjacency matrix strategies, produces mixed results. The spatial adjacency variant (MAP-ViGAT (F)) delivered an MAE of 2.27, highlighting its capacity to capture localized dependencies, such as adjacent regions with correlated intensity fluctuations. This approach is effective because these correlations often result from coherent and modal couplings. In contrast, the feature-based adjacency variant (MAP-ViGAT (S)) performed worse, with an MAE of 5.87, indicating that it struggled to capture the intricate modal interactions necessary for accurate temperature predictions. The learnable adjacency variant (MAP-ViGAT (L)) showed promise with an MAE of 3.35; however, further refinement was required to fully exploit the critical interference patterns.

The learnable importance-weighted non-symmetric attention vision transformer (LINA-ViT) introduced an interesting approach by assigning learnable importance weights to patches in the specklegram. This model achieved an MAE of 2.93, indicating its ability to prioritize areas that are sensitive to temperature changes. However, the challenge of balancing attention across different regions of a specklegram remains. Overemphasis on certain regions could lead to underrepresentation of other critical areas, resulting in suboptimal global interference pattern modeling.

**Traditional Models and Custom CNN**

The custom CNN performed reasonably well with an MAE of 2.98. While this model effectively extracted local spatial features from speckle patterns, it struggled with complex modal interactions and global phase shifts induced by temperature variations. The CNN architecture, which focused on localized feature extraction, failed to capture the broad interference patterns and phase shifts that are critical for accurate temperature estimation. Despite its limitations, the custom CNN outperformed standard architectures, such as MobileNet, InceptionNet V2, and ResNet, making it a reliable baseline model for specklegram analysis.

**Comparison With Literature**

Table 3. presents a comparison between the current study and other studies. Using a CNN–ANN hybrid model, Arango et al. [32] reported an MAE of 2.34 using a CNN-ANN hybrid model, which underscores the limitations of CNN-based approaches in handling the global interference effects present in specklegram data. This hybrid model struggled to capture broader interference patterns and phase shifts, resulting in higher MAE values compared to transformer-based architectures. Similarly, Vélez et al. [33] reported an MSE of 8.19 using an ANN, further indicating that traditional neural networks face challenges in modeling complex interactions in FSS data.

The results clearly demonstrate that transformer-based architectures, particularly the Vision Transformer (ViT) and Graph Attention Vision Transformer (GAT-ViT), are highly effective for temperature prediction using FSS data and have set a new benchmark in this domain. Their ability to capture both local intensity variations and global phase relationships as well as model complex interference patterns positions them as the best-suited models for this application. Although traditional CNNs and hybrid models have their merits, they fall short in handling the intricate modal interactions and phase shifts that are characteristic of specklegram data.

28# 7 Conclusion

This study demonstrates the transformative potential of transformer-based architectures, especially Vision Transformers (ViTs) and Graph Attention Vision Transformers (GAT-ViTs), for temperature prediction using Fiber Specklegram Sensing (FSS) data. These models capture complex relationships within high-dimensional specklegram patterns, significantly outperforming traditional models, such as CNNs and hybrid approaches. The superior performance of these transformer models is attributed to their ability to model both local intensity variations and global phase relationships, which are critical for understanding nonlinear interference patterns in optical fibers under temperature variations.

The success of transformer architectures in this context represents an important step forward in the application of deep learning to optical sensing technologies. By setting a new benchmark for temperature prediction accuracy, this study opens up exciting possibilities for their use in broader environmental monitoring and industrial applications, where precise temperature control and monitoring are essential.

However, challenges remain in the refinement of these methods. For instance, the performance of advanced models, such as Swin Transformers and LINA-ViT, suggests that architectural optimizations, such as incorporating better global context modeling and adaptive attention mechanisms, can lead to even more precise predictions. Future research could explore these enhancements along with real-time deployment strategies to fully exploit the potential of transformers in optical sensing systems.

A promising avenue for future work involves improving model interpretability and transparency using other advanced techniques. This would provide more insights into the decision-making process of these advanced models, thereby increasing their applicability in critical fields, such as structural health monitoring and industrial process control. Moreover, further investigation of hybrid models combining transformers with other advanced neural network architectures could lead to even greater performance gains in this emerging area of research.

# 8 References

[1] Al Zain, Mostafa, et al. "A high-sensitive fiber specklegram refractive index sensor with micro-fiber adjustable sensing area." IEEE Sensors Journal 23.14 (2023): 15570-15577.
[2] Inalegwu, O. C., Roman, M., Mumtaz, F., Zhang, B., Nambisan, A., & Huang, J. (2024, June). Optical fiber specklegram sensor for water leak detection and localization. In Optical Waveguide and Laser Sensors III (Vol. 13044, pp. 53-59). SPIE.
[3] Vélez-Hoyos, F. J., Aristizábal-Tique, V. H., Quijano-Pérez, J. C., Gómez-López, J. A., Trujillo, C., & Herrera-Ramirez, J. (2024). Fiber Speckle Sensing Scheme By Optical Power Filtering: Performance Analysis For Temperature Measurements.
[4] Li, G., Liu, Y., Qin, Q., Pang, L., Ren, W., Wei, J., & Wang, M. (2023). Fiber specklegram torsion sensor based on residual network. Optical Fiber Technology, 80, 103446.